\documentclass [letterpaper]{IEEEtran}
\usepackage{graphicx,
	psfrag,
	epsfig,
	epstopdf,
	amsthm,
	amssymb,
	url,
	subcaption,
	balance,
	enumerate,
	color,
	setspace,
	tikz,
	pgfplots
}
\usepackage{amsmath}
\usepackage[nospace,noadjust]{cite}
\usepackage{pbox}
\usepackage{multirow}
\usepackage{adjustbox}
\usepackage{array}
\usepackage{booktabs}
\usepackage{float}
\usepackage{algorithmic}
\usepackage{algorithm}
\usepackage{comment}
\usepackage{breqn}
\usetikzlibrary{shapes.geometric}

\newcommand{\cref}[1]{Constraint~\ref{#1}}

\newcommand{\ignore}[1]{}
\usepackage[paperwidth=8.5in, paperheight=11in,top=1.1in, bottom=0.99in, left=0.61in, right=0.61in]{geometry}

\usepackage{comment}

\begin{document}

\title{Distributed Traffic Control in Complex Dynamic Roadblocks: A Multi-Agent Deep RL Approach}
	\author{ 
	\IEEEauthorblockN{Noor Aboueleneen\IEEEauthorrefmark{1},  Yahuza Bello\IEEEauthorrefmark{2}, Abdullatif Albaseer\IEEEauthorrefmark{1}, Ahmed Refaey Hussein \IEEEauthorrefmark{2}, Mohamed Abdallah\IEEEauthorrefmark{1}, and Ekram Hossain  \IEEEauthorrefmark{3}}\\

        \IEEEauthorblockA{\IEEEauthorrefmark{1} College of Science and Engineering, Hamad Bin Khalifa University, Doha, Qatar}\\
        \IEEEauthorblockA{\IEEEauthorrefmark{2} University of Guelph, Guelph, Ontario, Canada}\\
	\IEEEauthorblockA{\IEEEauthorrefmark{3} University of Manitoba, Winnipeg, Manitoba, Canada}}

\maketitle
\begin{abstract}
Autonomous Vehicles (AVs) represent a transformative advancement in the transportation industry. These vehicles have sophisticated sensors, advanced algorithms, and powerful computing systems that allow them to navigate and operate without direct human intervention. However, AVs' systems still get overwhelmed when they encounter a complex dynamic change in the environment resulting from an accident or a roadblock for maintenance. The advanced features of Sixth Generation (6G) technology are set to offer strong support to AVs, enabling real-time data exchange and management of complex driving maneuvers. This paper proposes a Multi-Agent Reinforcement Learning (MARL) framework to improve AVs' decision-making in dynamic and complex Intelligent Transportation Systems (ITS) utilizing 6G-V2X communication. The primary objective is to enable AVs to avoid roadblocks efficiently by changing lanes while maintaining optimal traffic flow and maximizing the mean harmonic speed. To ensure realistic operations, key constraints such as minimum vehicle speed, roadblock count, and lane change frequency are integrated. We train and test the proposed MARL model with two traffic simulation scenarios using the SUMO and TraCI interface. Through extensive simulations, we demonstrate that the proposed model adapts to various traffic conditions and achieves efficient and robust traffic flow management. The trained model effectively navigates dynamic roadblocks, promoting improved traffic efficiency in AV operations with more than 70\%  efficiency over other benchmark solutions.
\end{abstract}


\section{Introduction}
 \IEEEPARstart{O}{ver} the last two decades, academics have focused on novel approaches that can improve security and provide comfort in Intelligent Transportation Systems (ITS) \cite{8771378}. Safety is a primary design objective for autonomous driving because it can enable vehicles to transverse roads independently with minimal human intervention. This autonomy aims to reduce accidents caused by human impairments such as sickness and fatigue. Current autonomous driving systems rely on an array of onboard sensors to collect pertinent inputs \cite{9617150,trackingsensors}. Sharing this input with various participating Autonomous Vehicles (AVs) can significantly improve the efficiency of ITS. The development of advanced communication technologies like Vehicle-to-Vehicle (V2V), Vehicle-to-Infrastructure (V2I), and Vehicle-to-Everything (V2X) facilitates efficient wireless communication within the ITS ecosystem \cite{ouaissa2022secure}. Specifically, 6G-V2X can facilitate fast and reliable real-time data sharing among cooperating AVs and between AVs and the infrastructure, improving their environmental awareness beyond the information collected by onboard sensors 


A crucial element for AVs navigating complex driving scenarios is a highly accurate decision-making control module that operates almost instantaneously. This module directs the vehicle's action execution system to execute various maneuvers such as car following, obstacle avoidance, lane changing, and overtaking \cite{rosenzweig2015review}. Among these driving maneuvers, car following and lane changing are the two most frequent and significant driving tasks \cite{k} \cite{l}. Consequently, a substantial body of research has focused on developing models for car following and lane changing \cite{9439506,9057371,9468359, wang2023car,wang2022velocity,an2020car,9810922,9837330,peng2022integrated, wang2021dynamic,wang2021intelligent,xia2021human,he2023modeling}. Nonetheless, the studies typically examine car-following and lane-changing behaviors in isolation. Many scholars have highlighted the impact of lane-changing on vehicles in adjacent lanes, emphasizing the need for integrated research that concurrently addresses both car-following and lane-changing behaviors. Such comprehensive studies on vehicle driving systems are paramount.

Lately, the car-following models that adopt Machine Learning (ML) have demonstrated remarkable performance, particularly those utilizing Reinforcement Learning (RL) and Deep RL (DRL) techniques \cite{9439506,9057371,9468359, wang2023car,wang2022velocity,an2020car}. Similarly, lane-changing models are benefiting from these advancements, with researchers increasingly utilizing RL and DRL methods like Deep Q-Network (DQN) and Deep Deterministic Policy Gradient (DDPG) algorithms to address the complexities of lane-changing maneuvers \cite{9810922,9837330,peng2022integrated, wang2021dynamic,wang2021intelligent,xia2021human,he2023modeling}. The integration of DRL algorithms to develop decision modules that can manage car following and lane changing actions has been explored in recent research \cite{zhang2018lane,10437453, 10622833, m, n, 8569568, makantasis2020deep, wang2021modeling, shi2022integrated}. However, only a handful of studies have investigated how the coordination of AVs can improve the performance of this combined approach. In response to this, Multi-Agent RL (MARL), which leverages centralized training and decentralized execution, offers a robust approach that can enable AVs to coordinate and execute complex driving maneuvers more realistically. Researchers in the domain of ITS, especially in traffic control, have begun to embrace MARL in designing coordinated traffic control frameworks \cite{8667868,10368025, calvo2018heterogeneous, o, p, van2016coordinated,8686046}. Nevertheless, the application of MARL in integrating both lane-changing and car-following maneuvers remains relatively unexplored in existing research.


To this end, we propose a decentralized multi-agent model utilizing DRL for autonomous lane-changing and car-following decision control in a dynamic roadblock setting utilizing the 6G-V2X communication. Specifically, the proposed MARL-based model aims to provide autonomous cars with the adequate decision of whether to change lanes or stay in the current lane based on the gathered information within the ITS environment. Most of the proposed approaches for optimal decision-making regarding lane-changing and car-following for AVs deploy a centralized methodology to train the models and execute driving maneuvers. This requires high computational capabilities to gather and transmit the data of various vehicles and transmit the decisions chosen to each vehicle within the environment. Furthermore, the ITS research domain focuses on controlling the traffic light systems, rather than the vehicles, to facilitate traffic flow and ensure safe routing for vehicles, based on the data acquired from the environment. Given the shortcomings mentioned above, this research's contributions are as follows:
\begin{itemize}
    \item We aim to improve traffic flow in a dynamic roadblock environment in an ITS by optimizing the AVs' decision-making process to ensure the maximization of the harmonic speed. The proposed model ensures smooth and efficient traffic flow, reducing congestion and improving overall traffic dynamics while guaranteeing that AVs can adapt their lane-changing strategies in real-time, maintaining optimal driving performance even in complex traffic scenarios.   
    \item We propose a decentralized multi-agent model that allows each AV to make independent lane-changing and car-following decisions within a multiple roadblock environment. This eliminates the need for a centralized system, thereby reducing computational overhead and improving the scalability of the solution. Each vehicle operates as an agent that uses local observations to make real-time decisions, promoting efficient and adaptive behavior in dynamic traffic environments.
    \item Our extensive simulations demonstrate significant improvements due to the proposed MARL algorithm in terms of various performance metrics such as convergence rate, stability under dynamic conditions, and overall traffic flow efficiency by more than 70\% over the other proposed solutions. 
\end{itemize}


The rest of the paper is organized as follows: Section II outlines the preliminaries and reviews relevant literature, which includes an overview of MARL and a discussion of its application in achieving efficient driving maneuvers. Section III details the proposed system model, formulates it as a Markov Decision Process (MDP) and describes the solution approach. Section IV provides the performance evaluation, and Section V  presents the concluding remarks.

\begin{table*}
\begin{center}
\caption{Summary of Different Models in terms of Traffic Control, Lane Changing, Car Following and Roadblocks Avoidance}\label{table:CPSLT}
\begin{tabular}{ | m{7em} | m{2em} | m{3em} | m{3em} |  m{5em} | m{6em} |  m{18em} | }
  \hline
Domain & Ref & Single-agent  & Multi-agent &  Coordination among agents & Algorithm  & Limitations \\
  \hline
  \multirow{3}{7em}{Car following models}  & \cite{9439506}& \hfil \checkmark & \hfil -  & \hfil - & DQN &  The proposed model may encounter computational inefficiencies because of video frame processing and thus will struggle to adapt to real-time scenarios\\ \cline{2-7}
                                         & \cite{9468359} & \hfil \checkmark & \hfil - & \hfil - & Adjusted Tsallis Actor-Critic (ATAC) & The proposed model utilizes NGSIM data to develop the training environment tailored to a specific scenario, which will limits its applicability to other road scenarios \\ \hline
                                           
   \multirow{3}{7em}{Lane changing models}  & \cite{9837330}& \hfil \checkmark & \hfil - & \hfil -  & Maximun entropy inverse RL & The proposed framework relies heavily on specific features and predefined scenarios for training. This can restrict the model's capability for generalization across various real-world scenarios \\ \cline{2-7}
                                           & \cite{peng2022integrated} & \hfil \checkmark & \hfil - & \hfil - & D3QN and DDPG & The study assumed that lane changing maneuvers were instantaneous for simplicity, which leads to the exclusion of the integration of detailed lane-change trajectories in the analysis process, limiting the study's ability to account for the complexities of real-world lane changes  \\  \hline

  \multirow{4}{7em}{Integrated car following and lane changing models}  & \cite{8638814} & \hfil -  & \hfil \checkmark & \hfil \checkmark & DQN-based & The dynamic coordination approach may face challenges scaling in highly complex traffic environments with numerous agents. Additionally, the frequent communication between agents to update coordination graphs can introduce unforeseen latency issues, affecting real-time applications \\ \cline{2-7}
                                           & \cite{10367764} & \hfil - & \hfil \checkmark & \hfil \checkmark & MA2C & The reliance on information sharing among agents can be impacted by communication delays or failures, potentially affecting decision-making quality and safety \\  \cline{2-7}
                                           & \cite{10437453} & \hfil \checkmark & \hfil - & \hfil - & DQN-based & The proposed MDP model is specifically designed for roadblock scenarios, and it does not account for the coordination between neighboring vehicles, limiting its practicality and effectiveness in real-world scenarios \\  \hline                                           

  \multirow{5}{7em}{Traffic control models} & \cite{8667868}& \hfil - & \hfil \checkmark & \hfil - & Multi-agent Advantage Actor-Critic (MA2C) & The robustness of the proposed MA2C can be compromised by noisy and delayed state measurements, which may impede its convergence in a large-scale traffic scenario \\ \cline{2-7}
                                           & \cite{10637352} & \hfil - & \hfil \checkmark & \hfil \checkmark & Novel RegionSTLight & The model's adaptability to various urban traffic conditions and unforeseen scenarios, such as sudden traffic surges, might be limited \\  \cline{2-7}
                                           &  \cite{10531693} & \hfil - & \hfil \checkmark & \hfil \checkmark & Novel Multi-Agent Adaptive Broad DRL (ABDRL) & The approach may encounter scalability issues when applied to larger, and more complex traffic networks in practical applications due to increased computational requirements \\

 \hline
\end{tabular}
\end{center}
\end{table*}

\section{Background and Related Works}

This section provides a review of the concept of MARL and the existing solution approaches found in the literature. Additionally, we review relevant research studies that employ MARL to develop efficient traffic control systems and driving maneuvers within the ITS domain.

\subsection{Multi-Agent Reinforcement Learning}

RL is formally modeled as an MDP, 
characterized by the tuple $<S, A, P, R, \gamma>$. Here, $S$ represents the state space, $A$ represents the action space, $P$ is the probability transition function, $R$ defines the reward function, and $\gamma$ is a discount factor that balances the immediate and future rewards. For every time step $t$ of the current state $s_t \in S$, the agent selects an action $a_t \in A$ according to defined policy $\pi(a_t|s_t)$, which maps state $s_t$ to its corresponding action $a_t$. For the action $a_t$ taken, the agent receives a reward $r_t \in R$ from the environment and transitions to the next state $s_{t+1}$ according to the $P$. The objective goal of the agent is to maximize the discounted expected reward from each state $s_t$. Numerous methodologies for identifying feasible solutions in the realm of RL have been extensively documented in the literature, ranging from value-based techniques to policy-based techniques \cite{wang2022deep}. 

MARL extends the concept of single-agent RL to scenarios involving multiple agents. Like single-agent RL, MARL tackles sequential decision-making problems, but the state evolution and individual rewards are determined by the collective actions of all agents involved. Formally, MARL is often modeled as a Markov game \cite{Zhang2021}, defined by the tuple $<N, S, A_{i\in N}, P, R_{i \in N}, \gamma>$, where $N$ denotes the set of agents, $S$ represents the state space, $A_{i\in N}$ signifies the set of actions of all agents, with $A_i$ being the action of agent $A_i$, $P$ is the probability transition function, $R_{i \in N}$ is the reward function providing the immediate reward for agent $i$ when transitioning from state $s_t$ to the next state $s_{t+1}$, and $\gamma$ is the discount factor.

Various solution approaches for MARL have been proposed in the literature, including Independent Q-learning (IQL), Independent Advantage Actor-Critic (IA2C), Multi-agent Independent Advantage Actor-Critic (MA2C), and Multi-Agent Deep Deterministic Policy Gradients (MADDPG). The most widely adopted is the IQL algorithm, where each agent trains independently and updates its policy dynamically. The main challenge with this approach is achieving global convergence, as each agent considers its counterparts part of the environment, resulting in a partially observable and nonstationary environment. To address this issue, researchers have proposed multiple techniques, with recent trends favoring methods that utilize parameter sharing among agents, such as MAPPO, MA2C, and MADDPG. For this work, the MADDPG algorithm is used to train the agents and develop effective driving maneuvers to avoid roadblocks \cite {q}.


\subsection{State-of-the-Art  Models for Driving Maneuvers}

\subsubsection{Traffic control models} 
Numerous research efforts have concentrated on utilizing MARL to design control modules for effective traffic management and driving maneuvers. For example, in \cite{8667868}, the authors extended the widely known IQL algorithm to MA2C approach specifically for Traffic Signal Control (TSC). Additionally, \cite{10368025} proposed a framework for coordinated Adaptive TSC using a pressure indicator that captures traffic conditions across a large-scale intersection network. The authors developed a reward function incorporating this pressure indicator and vehicle waiting times to enable efficient coordination among multiple intersections. \cite{calvo2018heterogeneous} presented a solution for a heterogeneous multi-junction Urban Traffic Control (UTC) scenario. The authors utilize the Independent DQN (IDQN) to enable learning across multiple agents, where each agent employs a Dueling Double DQN (DDDQN) with a fingerprint to address the non-stationarity inherent in multi-agent environments. \cite{10637352} studies city-wide traffic signal control at multiple intersections. The authors introduce a novel RegionSTLight algorithm that utilizes a regional MARL framework to break down the global traffic control problem into smaller regional zones based on real-time traffic density. In \cite{10531693}, the authors proposed a Multi-Agent Adaptive Broad
DRL (ABDRL) framework for TSC, which combines a broad network to capture diverse feature interactions with a deep network for complex pattern recognition and decision-making.

\begin{figure*}[ht!]
\centering
   \includegraphics[width=\textwidth]{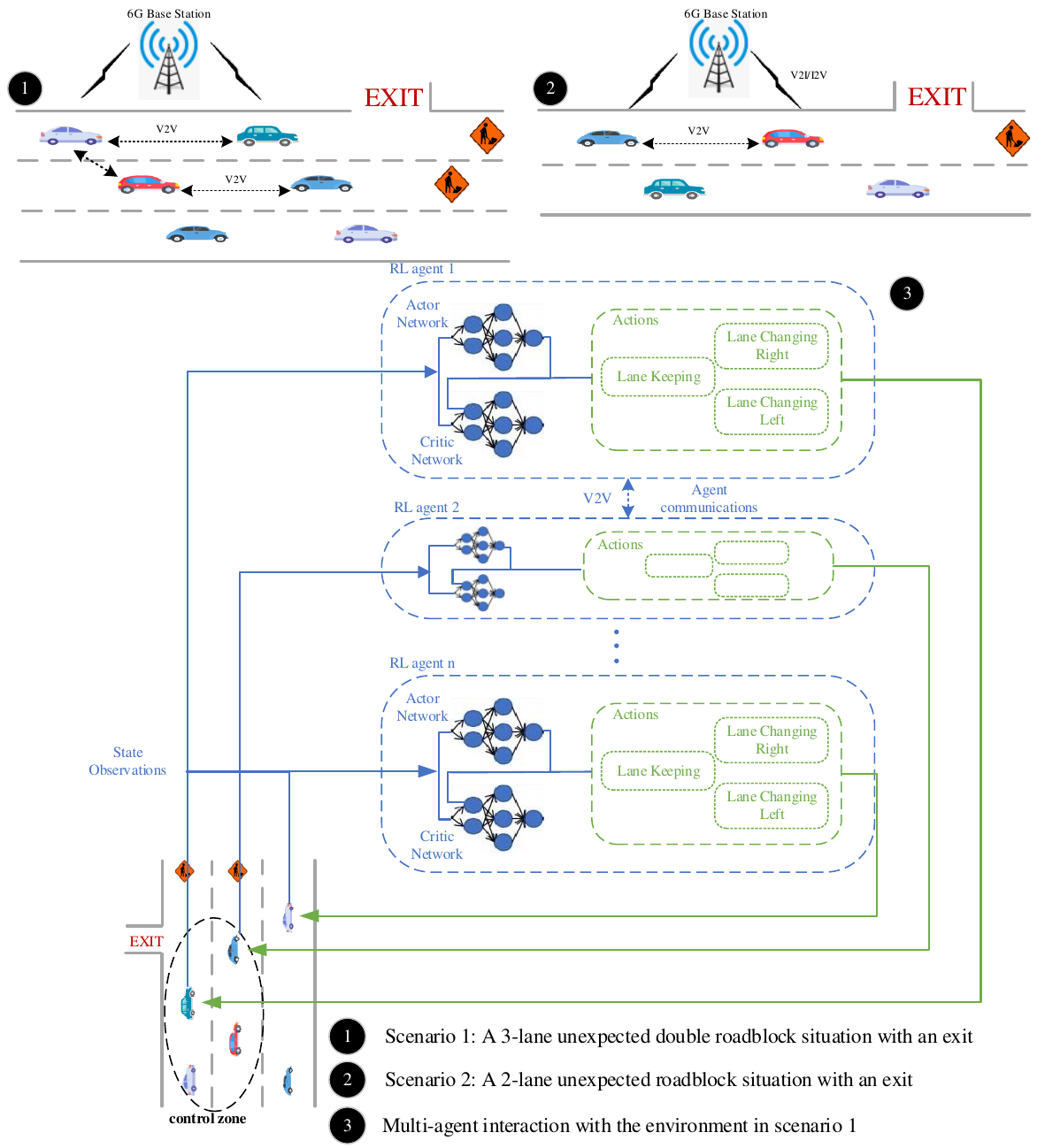}
    \caption{MARL-based roadblock control system and highway roadblock scenarios}
    \label{fig:Fig1}
\end{figure*}

\subsubsection{Car-following and lane-changing models}
Significant research efforts over the past decade have focused on developing car-following and lane-changing models. In \cite{9439506}, the authors propose a car-following framework for autonomous vehicles (AVs) using Red Green Blue Depth (RGB-D) frames for navigation. The system integrates the You Look Once version 3 (YOLOv3) object detection model to recognize lead vehicles and obstacles, while RL algorithms are employed to make real-time driving decisions. In \cite{9468359}, a novel car-following model with a longitudinal control strategy for level 3 autonomous driving that adopts the Tsallis actor-critic (ATAC) algorithm with automatic entropy adjustment is proposed. The authors utilize 1,641 pairs of car-following trajectories derived from the Next Generation Simulation (NGSIM) dataset to train an RL agent, ensuring the development of an efficient car-following model. 

The authors in \cite{9837330} developed a segmented lane-change trajectory planning method for AVs on closed highways. Their proposal incorporates the Maximum Entropy Inverse RL (IRL) algorithm to learn from human driving behavior and the Long Short-Term Memory (LSTM) and Bayesian Network (BN)-based model to predict the driving intentions of target vehicles. Peng et al. \cite{peng2022integrated} addressed the issues of poor decision-making coupling,  high scene complexity, and safety assurance by focusing on the priorities and logic of lane changing and car following decisions, taking into account comfort, efficiency, and safety. A double-layer decision-making model is constructed, utilizing two DRL algorithms (D3QN and DDPG) to handle large-scale mixed-state spaces and ensure effective lane-changing and car-following actions. Their analysis showed their proposed approach's effectiveness compared to other techniques. In addition, the authors in \cite{zhang2022multi} introduced a bi-level lane-changing planning maneuver to address the issue of traditional lane change decision models that often treat it as a one-agent problem, neglecting the essential multi-agent interactions in traffic, which can result in hazardous behaviors and reduced traffic efficiency. The authors proposed an upper level with a MADRL-based decision model, while the lower level includes a negotiation-based right-of-way assignment model. An extensive simulation showed the efficiency of the authors' proposed technique in their research. The authors in \cite{li2023safe} proposed a Risk-fused Constraint DRL (RCDRL) approach using a D3QN network for safer lane change decisions to tackle the safety issues associated with DRL when deployed for autonomous driving. This method effectively handled unsafe decision actions during policy network training. After that, the RCDRL policy was evaluated under various traffic conditions, including real vehicle tests, showing superior performance in safe decision-making. However, the authors in their work only focused on the safety aspect of AVs lane-changing and did not take into consideration the scalability aspect, which studies the model with an increasing number of vehicles and traffic complexities, which is crucial for ITS environments.

\subsubsection{Integrated car-following and lane-changing models}

Recently, a few proposals have emerged proposing a combined car-following and lane-changing models. For example, the authors in \cite{10437453} proposed an integrated decision-control system for car-following and lane-changing using DRL. The proposed framework is tailored to handle roadblock scenarios such as the occurrence of construction sites on highways. The authors model the roadblock scenario as an MDP and utilize the DQN algorithm to train an agent to make appropriate decisions, such as staying in the current lane or switching lanes, based on real-time traffic conditions. In \cite{8638814}, the authors introduced a dynamic coordination graph to model the continuous topology changes of interacting AVs on a highway. They present a DQN-based algorithm that enables effective driving maneuvers in this dynamic environment. The authors in \cite{chen2023deep} addressed an on-ramp merging challenge in a highway scenario by framing it as a MARL problem. AVs that are merging or staying in their lane work together to obtain a strategy that adapts to human-driven vehicles to maximize overall traffic throughput. In \cite{10367764}, a MARL framework for Connected Autonomous Vehicles (CAVs) that utilizes information sharing to allow AVs to perform safe driving maneuvers is proposed.

Table \ref{table:CPSLT} provides an overview of recent studies that utilized DRL in the context of road navigation, detailing whether each study adopts single or multiple agents, the level of cooperation among agents in decision-making, the specific algorithms used or proposed, and any identified limitations of these approaches. Our approach stands out from existing proposals in the literature in two significant ways. Firstly, we tailor our framework to a roadblock scenario on a highway, where the agents are AVs navigating the road, unlike most traffic control studies that consider the traffic control systems as agents. Secondly, we employ the MADDPG algorithm to identify effective driving maneuvers, whereas most studies typically use the IQL algorithm for determining driving actions.

 
\begin{table}[ht!]
\begin{center}
\caption{Main Notations}\label{table:symbol}
\begin{tabular}{ | m{5em} | m{18em}  |}
  \hline
 Symbols & Description  \\
  \hline
$M$ & Total number of vehicles    \\ 
  \hline
 $v_i(t)$   & Speed of $i$-th vehicle    \\ 
 \hline
  $l_i(t)$   & lane of $i$-th vehicle    \\ 
 \hline
  $V_{min}$  &  Minimum required vehicle speed  \\   
 \hline
  $R_i$  &   Number of roadblocks with $i \in \{1,2\}$  \\ 
 \hline
   $f_{lc}$ & Frequency of lane-change    \\  
  \hline
   $f_{max}$ & Maximum number of lane-change    \\ 
    \hline
   $x_i(t)$, $y_i(t)$ & Location of the $i$-th vehicle   \\ 
   \hline
   $a_i(t)$ & acceleration of the $i$-th vehicle   \\ 
   \hline
   $d_{i2j}(t)$ & Distance of the $i$-th vehicle to the $j$-th vehicle     \\ 
   \hline
   $d_{i2R}(t)$ &  Distance of the $i$-th vehicle to the roadblock   \\ 
   \hline
   $d_{i2E}(t)$ &  Distance of the $i$-th vehicle to the exit   \\ 
   \hline
   $A_{lk}(t)$ & Action of lane-keeping    \\ 
   \hline
   $A_{lc}(t)$ &  Action of lane-chaanging   \\ 
  \hline
   $V_{h}(t)$ &  Harmonic speed   \\ 
 \hline
\end{tabular}
\end{center}
\end{table}

\section{System Model and Assumptions}

This section describes the proposed roadblock control model and the scenarios considered. Next, we formalize the MDP model. Finally, we discuss our solution approach that adopts the MADDPG algorithm to find effective driving maneuvers.

\subsection{Proposed Model and Driving Scenario}

The proposed roadblock control model is illustrated in Fig.~\ref{fig:Fig1}. We envision a highway scenario and  
to model these scenarios accurately, we assume that AVs have the necessary sensors and advanced 6G-V2X communication systems, enabling them to communicate with nearby vehicles and the nearby 6G base station in case the processing capabilities of the AVs become overwhelmed. This setup allows vehicles to gather critical information, including velocity, acceleration, AVs' locations, distance to various roadblocks, and other similar data about participating AVs.

As shown in Fig.\ref{fig:Fig1}, two scenarios are considered in our implementation. First, a three-lane highway with two roadblocks ahead of the AVs and an optional exit, where the vehicles must choose to change lanes, stay in the lane, or take the exit. Second, a two-lane highway with one roadblock and an optional exit, where the AVs apply the decision taken to avoid the roadblock with minimal congestion. The roadblocks are considered to be accidents or traffic maintenance in the proposed model. AVs are considered the agents of the MARL in the proposed model. As an agent, each AV will gather real-time data from its surroundings, process this information using the MARL algorithm, and execute optimal decisions to enhance traffic flow and safety. This approach aims to ensure efficient navigation and reduce congestion by leveraging DRL techniques. AVs communicate their decisions within the environment to establish the optimal traffic flow with minimal hazards. This communication ensures that each vehicle's actions are coordinated, allowing for smoother lane changes, reduced congestion, and enhanced safety. By sharing real-time data about road conditions, obstacles, and traffic flow, the vehicles can obtain sound decisions that collectively optimize the traffic system, leading to a more efficient and secure driving experience.

In our model, AVs assess the state information acquired by their equipped sensors. These sensors retrieve real-time data about the environment, creating comprehensive situational awareness. After processing this data through our MARL proposed model, the vehicles make decisions on whether to change lanes or remain in their current lane. This decision-making process is influenced by multiple input factors that are critical to ensuring efficiency and smooth traffic flow. AV's speed is one of the primary factors considered. The autonomous system continuously monitors the speed of the AV and the surrounding traffic. This includes the relative speed of vehicles in adjacent lanes to determine if a lane change would be beneficial or necessary. Maintaining an optimal speed relative to other vehicles helps in making safe and efficient lane changes. The system also calculates the distance between vehicles. It assesses the distance to the vehicles ahead and in adjacent lanes to maintain a safe following distance. This is crucial for ensuring smooth braking and acceleration. The system can make more informed decisions about when to change lanes by keeping a safe buffer zone around the vehicle.

Another key factor is the distance to roadblocks. The sensors detect any obstacles or roadblocks, such as stalled vehicles, debris, or construction zones. The vehicle must decide whether a lane change is required to avoid these hazards. Ensuring that the vehicle can safely maneuver around obstacles is essential for maintaining continuous and safe travel. Acceleration and deceleration are also assessed. The system evaluates the current rates of acceleration or deceleration to predict the future positions of the vehicle and the surrounding traffic. This helps in making timely decisions to change lanes or adjust speed, ensuring that the vehicle can respond quickly to dynamic traffic conditions. Finally, the location of vehicles is also part of the model's decision-making procedure.

By integrating these factors, the AV aims to make informed and efficient decisions regarding lane changes. Our model relies on a sophisticated sensor suite and advanced processing algorithms to evaluate real-time conditions and make dynamic lane change decisions. This ensures that AVs operate efficiently, adapting to the complexities of real-world driving environments.

In our problem, the primary goal is to enable AVs to avoid roadblocks by changing lanes while maintaining maximum mean harmonic speed within the environment. This is accomplished by ensuring each AV does not exceed a reasonable number of lane changes. The mean harmonic speed is particularly useful in traffic flow analysis because it gives more weight to slower speeds, which often have a disproportionate impact on overall traffic efficiency and it is defined as follows:
\begin{equation}
    {V_h(t)= \sum_{t=1}^{T}  \frac{M}{\sum_{i=1}^{M} \frac{1}{v_i(t)}}}
\end{equation}
where $M$ is the total number of vehicles, 
$v_i$ is the speed of the $i$-th vehicle. 
By focusing on the harmonic mean, we can better understand and mitigate the effects of congestion and slower vehicles on the entire traffic system. 

In our study, the primary objective is to establish an efficient traffic flow that enables AVs to navigate around multiple dynamic roadblocks while maintaining the maximum achievable mean harmonic speed by changing lanes. Formally, our problem is formulated as follows:
\begin{equation}\label{1a}
    \textbf{P1}: \max_{ l_i(t) } \quad {V_h(t)},
\end{equation}

 \begin{equation}\label{1b}
  \textrm{s.t.} \quad  C1:  \\  V_i(t)\geq \ V_{min} , \forall i=\!\!1 \in M, \\
  \end{equation}
\begin{equation}\label{2b}
  \quad  C2:  \\  R_i\in \ \{1,2\},  
 \end{equation}
 \begin{equation} \label{3b}
    C3: f_{lc} \leq f_{max}, 
 \end{equation}
where $C$1, ensures the minimum required vehicle speed, constraint $C$2, indicates the number of dynamic roadblocks within the environment, while constraint $C$3 limits the frequency of lane changes by each autonomous vehicle to $f_{max}$, preventing excessive lane changes that can lead to traffic instability. Table \ref{table:symbol} summarizes the notations and their corresponding descriptions used in the formulation of the problem.

\subsection{Roadblock Control Problem as an MDP}
We formulate the AVs' decision-making procedure as an MDP model, where the AVs interact with each other and the environment. The MDP model lays out a mathematical framework that models decision-making in events where results are partly random and partly under the control of a decision-maker. The tuples of the MDP model are defined as follows.

\subsubsection{State Definition}
The state represents the current situation of the vehicle, including its position $(x_i(t),y_i(t))$, speed $v_i(t)$, acceleration $a_i(t)$, distance to other vehicles $d_{i2j}(t)$, distance to roadblocks $d_{i2R}$, and distance to exit lane $d_{i2E}$. The state captures all the necessary information required to make an informed decision about lane changes or lane keeping for the vehicle.
The state space of each agent at each time slot is defined as:
\begin{equation}
\begin{aligned}
  \mathbb{S}(t):  \langle  d_{i2j}(t), d_{i2E}(t), d_{i2R}(t), a_i(t),v_i(t),x_i(t),y_i(t) \rangle.
  \end{aligned}
\end{equation}
\subsubsection{Action Definition}
The actions are the set of all possible maneuvers the vehicle can take, such as changing lanes to the left, changing lanes to the right, or staying in the current lane. These actions are chosen based on the current state and aim to optimize the mean harmonic speed of all vehicles within the system. The action space of each agent at each time slot is defined as:

\begin{equation}
\begin{aligned}
\mathbb{A}(t)= \langle A_{lk}(t),A_{lc}(t)\rangle,
  \end{aligned}
\end{equation}
where $A_{lk}$ denotes the action of staying in the same current lane and $A_{lc}$ represents the action of changing the current lane to void the dynamic roadblock.

\subsubsection{Reward Definition}
The reward function assigns a numerical value to each state-action pair, representing the immediate benefit or cost of performing an action in a given state. The rewards are designed to encourage behaviors that maximize harmonic speed and minimize unnecessary lane changes. 
The reward function at each time slot is given as follows:

\begin{equation}\label{eqReward}
        \mathbb{R}(t) = \begin{cases}
            r1 + V_h(t) \\
            r2 + V_h(t) \\
            r3 + V_h(t) \\
            r4 + V_h(t)
        \end{cases}
    \end{equation}
where $r1$ represents a positive reward when the agent successfully avoids roadblocks by changing lanes or staying in the current lane without affecting the harmonic mean speed, $V_h$, and without violating any constraints, $r2$ represents a less positive reward for vehicles that managed to execute the chosen action, but they failed to merge smoothly into traffic. It is worth noting that $r2$ shows the action impacts on the speed but still adheres to the constraints. The last positive reward is $r3$ given to the agent when it takes the exit lane decision. Finally, $r4$ is a negative reward given to the vehicle as a violation penalty when it fails to merge with the traffic flow and encounters a roadblock, violates the minimum speed constraint $C1$, or exceeds the maximum number of lane changes $C3$. Note that these rewards are scalar values that are assigned appropriately based on the actions taken in the simulation phase.
    
\subsection{Solution Approach}
AV \(i\) receives the adjacent vehicle's action space and state space through a communication topology. In this topology, each agent's network shares this information with the following vehicle to ensure that the globally optimal policy within the network is generated. Communication topology refers to the structured way in which information is exchanged between agents. In this case, each autonomous vehicle communicates its state (such as position, speed, and acceleration) and potential actions to its neighboring vehicles. This exchange allows the vehicles to have a broader view of the environment and the actions of their peers, which is crucial for making informed decisions. By leveraging this shared information, each vehicle can make more accurate predictions about the future states of adjacent vehicles and adjust its actions accordingly, ensuring compliance with constraints such as maintaining minimum speed and limiting lane changes. This interconnected network of vehicles works collaboratively to optimize overall traffic flow while respecting all constraints. Communication ensures that decisions are not made in isolation but that the collective state and action possibilities of the surrounding vehicles are considered, promoting adherence to the above constraints

The goal of this model is to facilitate the generation of a globally optimal policy that satisfies all constraints. Instead of each vehicle solely focusing on its immediate rewards and states, the vehicles aim to maximize the collective benefit while ensuring that actions do not violate any constraints. This collaborative approach helps in avoiding conflicts and ensuring smoother transitions, especially in complex scenarios like lane changes and obstacle avoidance, without exceeding the maximum frequency of lane changes or dropping below the minimum speed. 

In our proposed approach, the agent's probability function to establish the optimal policy for an action $a$ in state $s$ is denoted by $\pi_i^*(a|s)$. This function represents the probability of executing action $a(t)$ in a given state $s(t)$ following the optimal policy $\pi^*$. The input to each agent's neural network is the current state $s(t)$, which includes information about the environment relevant to all agents. This state comprises the speed and position of the agent, distances to nearby vehicles, and information about roadblocks. In addition, the state space also includes information about the states of other agents that is observable and relevant to the decision-making process. The output provided by each agent's network is an action space representing the set of possible actions.  The network outputs these actions as a probability distribution, where each action $a_i$ is assigned a probability $\pi_i^*(a_i|s)$.

As illustrated in Algorithm \ref{a1g:three} and shown in Fig.\ref{fig:Fig1}, the MADDPG model is structured on the actor-critic methodology, where each agent has two neural networks, which are the actor-network $ \alpha (s ,a|\theta^\alpha)$  and the critic-network $ C (s ,a|\theta^C)$.  The actor-network is responsible for selecting actions. Given the current state $s(t)$, the actor-network outputs the action $a(t)$:
\begin{equation}
a(t) = \alpha(s(t)|\theta^\alpha).
\end{equation}
The critic-network evaluates the action taken by the actor-network by estimating the Q-value, which represents the expected cumulative reward. The Q-value is a function of the current state $s(t)$ and action $a(t)$ given by:
\begin{equation}
Q(s(t), a(t)|\theta^C) = C(s(t), a(t)|\theta^C).
\end{equation}
The actor and critic-networks within each agent iteratively update the parameters of the policy network $\theta^\alpha$ and the value network $\theta^C$, respectively. The policy update for the actor-network is given by the gradient ascent on the expected return as follows:
\begin{equation}
\nabla_{\theta^\alpha} J(\theta^\alpha) = \mathbb{E}_{s \sim \rho^\pi} \left[ \nabla_a Q(s, a|\theta^C) \nabla_{\theta^\alpha} \alpha(s|\theta^\alpha) \right].
\end{equation}
The critic-network is updated by minimizing the loss function, which is the mean squared error between the estimated Q-value and the target Q-value. The target Q-value $y(t)$ is given by:
\begin{equation}
y(t) = r(t) + \gamma Q(s(t+1), \alpha(s(t+1)|\theta^\alpha)|\theta^C),
\end{equation}
where $\gamma$ is the discount factor. The loss function $L(\theta^C)$ for the critic-network is:
\begin{equation}
L(\theta^C) = \mathbb{E}_{(s, a, r, s')} \left[ \left( y(t) - Q(s, a|\theta^C) \right)^2 \right].
\end{equation}
Simultaneously, the actor-network selects the immediate action $a(t)$ based on the current state $s(t)$, interacts with the environment to determine the next state $s_{t+1}$, and receives the current reward $r_t$ for the action taken:
\begin{equation}
s_{t+1} = \mathcal{E}(s(t), a(t))
\end{equation}
\begin{equation}
r_t = \mathcal{R}(s(t), a(t)),
\end{equation}
where $\mathcal{E}$ represents the environment transition function and $\mathcal{R}$ represents the reward function.

In this setup, the continuous interaction between the actor-network and the environment, along with the iterative updates of both the actor and critic-networks, guarantees that each agent learns to optimize its actions toward achieving the globally optimal policy.
Furthermore, the algorithm is capable of saving the transitions \( [s(t), a(t), r(t), s(t+1)] \) into the experience buffer \( B \) in order to allow the AVs to learn from the transitions of each other. Concurrently, the neural network executes the objective function \( J(\theta^u) \) and transmits it to the optimizer. This objective function is given as follows: 
\begin{equation}
     J(\theta^u) = \frac{1}{Z} \sum_{i=1}^{Z} (Q(s_i, a_i | \theta^Q) - y_i)^2. 
\end{equation}

Here, $Z$  represents the number of transitions stored in the experience buffer, \( Q(s_i, a_i | \theta^Q) \) denotes the Q-value estimated by the critic-network for the state-action pair \((s_i, a_i)\), while \( y_i \) represents the target value for this pair. The objective of the optimizer is to minimize the discrepancy between the estimated Q-value and the target value across all transitions stored in the experience buffer.

The actor's target network is crucial in selecting the next optimal action \(a(t+1)\). This selection is dependent on observations of the subsequent state \(s(t+1)\) acquired from sampling the transition buffer. Additionally, the soft update implementation of the network variables ensures a smooth transition between the online and target networks. This process improves the stability and convergence of the learning procedure. Mathematically, the actor's target network selects the action \(a(t+1)\) according to:
\begin{equation}
    a(t+1) = \alpha'(s(t+1)|\theta^{\alpha'}),
\end{equation}
where \(\alpha'\) represents the target actor-network.

Consequently, the critic's online network iteratively updates the value network variables \(\theta^{\theta^C}\), which represent the current $\theta^C$ value estimates. These updates are crucial for refining the critic's understanding of the environment and the actions taken by the agent (i.e., AV). The $\theta^C$ value estimation is given by:
\begin{equation}
    {\theta^C} (s, a | \theta^{\theta^C)} = C(s, a | \theta^{\theta^C}),
\end{equation}
where \(C\) denotes the critic-network.
Subsequently, the network constructs the loss function, quantifying the disparity between the estimated $\theta^C$ values and their corresponding target values. This loss function, denoted as \(L(\theta^\theta{}^C)\), is transmitted to the optimizer for parameter updates. The loss function is typically defined as the mean squared error MSE between the estimated $\theta^C$ values and their targets:
\begin{equation}
    L(\theta^{\theta^C}) = \frac{1}{Z} \sum_{i=1}^{Z} ({\theta^C}(s_i, a_i | \theta^{\theta^C}) - y_i)^2,
\end{equation}
where \(s_i\) and \(a_i\) are the states and actions from the transitions, and \(y_i\) denotes the target value for the \(i\)-th transition.
In summary, the actor's target network selects actions according to the future state observations, and the critic's online network updates \(\theta^C\) value estimates and computes the loss function, which guides the parameter updates through the optimizer. This iterative process enables the model to learn and improve its performance over time.

\begin{algorithm}[ht]
  \caption{Proposed MARL algorithm for solving $P_1$}
  {$M$, $V_{max}$, $L$,$R_0$,$R_1$}
\begin{algorithmic}[1]\label{a1g:three}

\STATE{Initialize weights $W$}
\STATE{Initialize each agent's actor $ \alpha (s ,a|\theta^\alpha)$ neural networks with weights $\theta^\alpha$}
\STATE{Initialize critic $ C (s ,a|\theta^C)$ neural network with weights $\theta^C$ }
\STATE{Initialize each agent's actor and critic target networks with weights $\theta^\alpha \gets \theta^{\alpha'}$ and $\theta^C \gets \theta^{C'}$, respectively  }
\STATE{Initialize experience buffer $b$ to capacity $\mu$}
  \FOR{$\text{episode } e=1 \text{ to } e=E $}
  \STATE{Get primary observations for roadblocks}
  \STATE{Initialize a random procedure $\delta$ for exploration of action}
    \STATE{ Initialize sequence $s_1$  }
      \FOR{$ \text{step } t = 1 \text{ to } t=T$ }
       \FOR{$ \text{agent } m = 1 \text{ to } m=M$ } 
      \STATE {For each agent $M$, select action $a_m(t)$ with respect to current policy and exploration noise under state $s_m(t)$ }
      \STATE{Execute actions of agents  $a(t)=(a_1(t),\dots,a_M(t)$ and get reward $r(t)$, enter new states $s_m(t+1)$ for each agent}
      \STATE{Store transition $[s(t), a(t), r(t), s(t+1)]$ in experience buffer $b$}
      \STATE{Sample random mini-batch of $Z$ transitions $(s(x), a(x), r(x), s(x+1))$ from $b$}
      \STATE{ Update replay buffer $b$ and store transition $[s(t), a(t), r(t), s(t+1)]$ }
      \STATE{Set $y_x=r_x+\gamma C '(s_{x+1} | \theta^{\alpha '}) |\theta^{C'} )$}
      \STATE{Minimize the loss to update the critic:}
      \STATE{$L= \frac{1}{Z} \sum_{x}(y_x - C (s_{x} ,a_{x},a_{x+1},\dots,a_X |\theta^C))^2$ }
      \STATE{Update policy of the actor using policy gradient}
      \STATE Update the target networks: 
      \STATE{$\theta^{\alpha '}\gets v\theta^{\alpha}+(1-v)\theta^{\alpha '}$} 
      \STATE{$\theta^{C'} \gets v\theta^{C}+(1-v)\theta^{C'}$}
      \ENDFOR
       \ENDFOR
        \ENDFOR
  \end{algorithmic}
\end{algorithm}

 \begin{figure}[ht]
    \centering
    \includegraphics[width=\columnwidth]{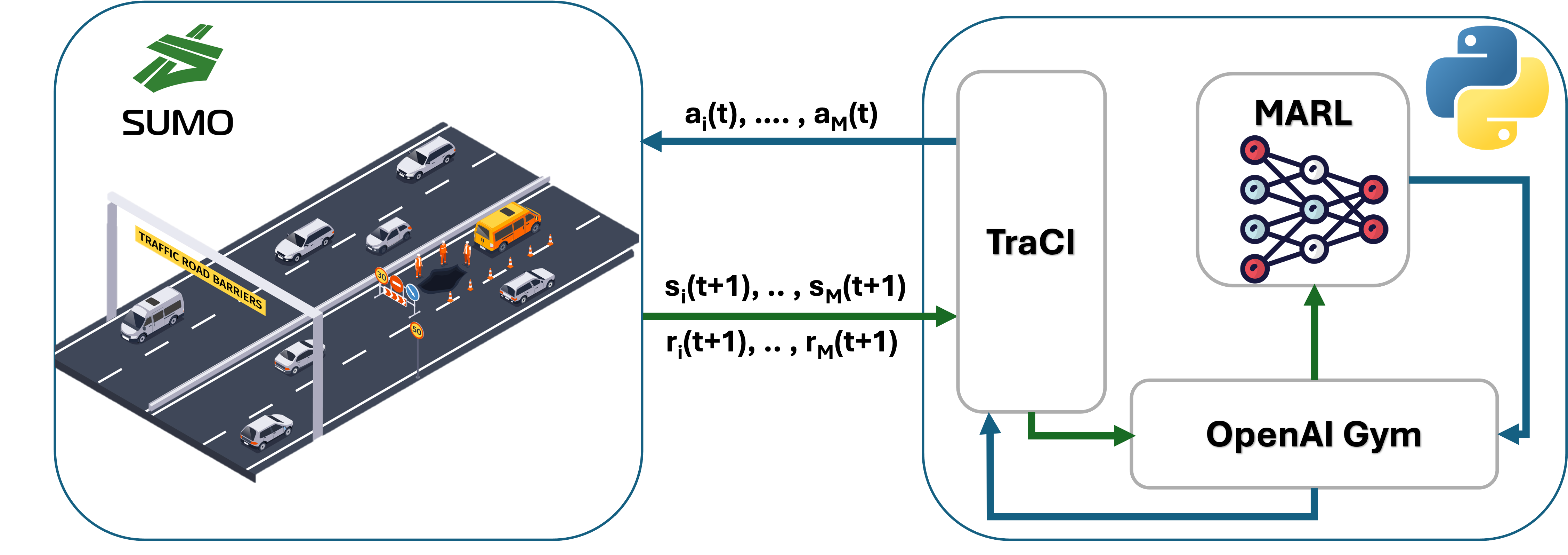}
    \caption{Illustration of the implementation environment using SUMO }
    \label{fig:e}
\end{figure}

\section{Performance Evaluation}
In this section, we provide an in-depth evaluation of the proposed solution, followed by a detailed analysis of the results.
\subsection{Experimental Setup} 
We utilized SUMO software with the TracI interface from Python to simulate traffic flow, thereby improving the authenticity and practicality of the proposed model. The TracI interface was specifically employed to control the lane-changing decisions for each agent in real-time while maintaining all constraints, allowing for a high degree of interaction and dynamism in the simulation. The model was trained for two distinct scenarios within an ITS environment. The first scenario involves a 300-meter, three-lane highway featuring two roadblocks and an exit lane. This setup was designed to test the model's ability to handle multiple obstacles and lane transitions effectively. The second scenario also involves a 300-meter highway but with two lanes, one roadblock, and an exit lane, as illustrated in Fig. \ref{fig:e}. This configuration presents a different level of complexity by reducing the number of lanes and roadblocks, thereby testing the model's adaptability to varying traffic conditions. To further challenge the model and ensure its robustness, the positions of the roadblocks within the environment were dynamically altered every 10 episodes. This periodic relocation of obstacles was intended to simulate real-world conditions where roadblocks and other traffic impediments can change location frequently, requiring the AVs to continuously adapt their lane changing and navigation strategies.

The MARL model is implemented using PyTorch, leveraging its robust framework for developing and training deep learning models. The simulation parameters are as follows: the MARL model was trained over 200 episodes, each containing 1000 epochs. The learning rate for the proposed MARL model was $1e^{-3}$, with a discount factor of 0.99, a batch size of 2096, and a memory capacity of $1e^6$. We set the values of $r_1$, $r_2$, $r_3$, and $r_4$ as 2, 1, 0, and -2, respectively. 
Within the simulation environment, the positions of the roadblocks were dynamically relocated after a specified period to various locations to improve the robustness of the training. A total of 4 AVs were utilized in the ITS during the training phase to ensure comprehensive interaction and learning. This setup allowed us to rigorously test and evaluate the performance of the MARL model in dynamically changing environments, simulating real-world conditions where roadblocks and traffic patterns can vary. The detailed results and insights derived from these simulations are discussed in the subsequent sections. After the training phase of the proposed model was completed, the agents' networks with their post-trained weights were saved for the testing phase. This ensured that the learned policies and strategies could be evaluated under controlled conditions, providing a clear understanding of the model's effectiveness and adaptation to real-world applications.

\subsection{Numerical Results} 
The model was tested with various numbers of vehicles, as illustrated in Fig. \ref{fig:Fig1}. This evaluation was designed to assess the efficacy of the proposed model in a highly dynamic environment. Specifically, the model was tested with 2, 4, and 6 vehicles to examine its adaptability and performance under different traffic densities.
The results demonstrated that the network could adapt to different environmental settings and achieve convergence across all tested scenarios. With just two vehicles, the model rapidly learned optimal navigation and lane-changing policies, demonstrating efficient roadblock avoidance and maintaining smooth traffic flow, all while adhering to the problem's defined constraints. When tested with 4 vehicles, the complexity increased, but the model maintained robust performance, effectively coordinating actions of whether to change lanes or stay in the same lane among the vehicles to optimize the overall traffic dynamics and ensure the avoidance of the roadblock. Testing with 6 vehicles presented the highest level of complexity, simulating a more congested environment. Despite this, the model was successfully adapted, demonstrating its scalability and robustness in handling increased traffic density with the existence of single or multiple roadblocks while following the problem's constraints. The adaptive capability of the model was further validated by its consistent performance in maintaining maximum harmonic speed while minimizing unnecessary lane changes. This highlights the model's effectiveness in achieving the desired balance of efficiency in an ITS environment. 

\begin{figure}
  \centering
  \captionsetup{belowskip=0pt}
  \includegraphics[width=0.9\linewidth]{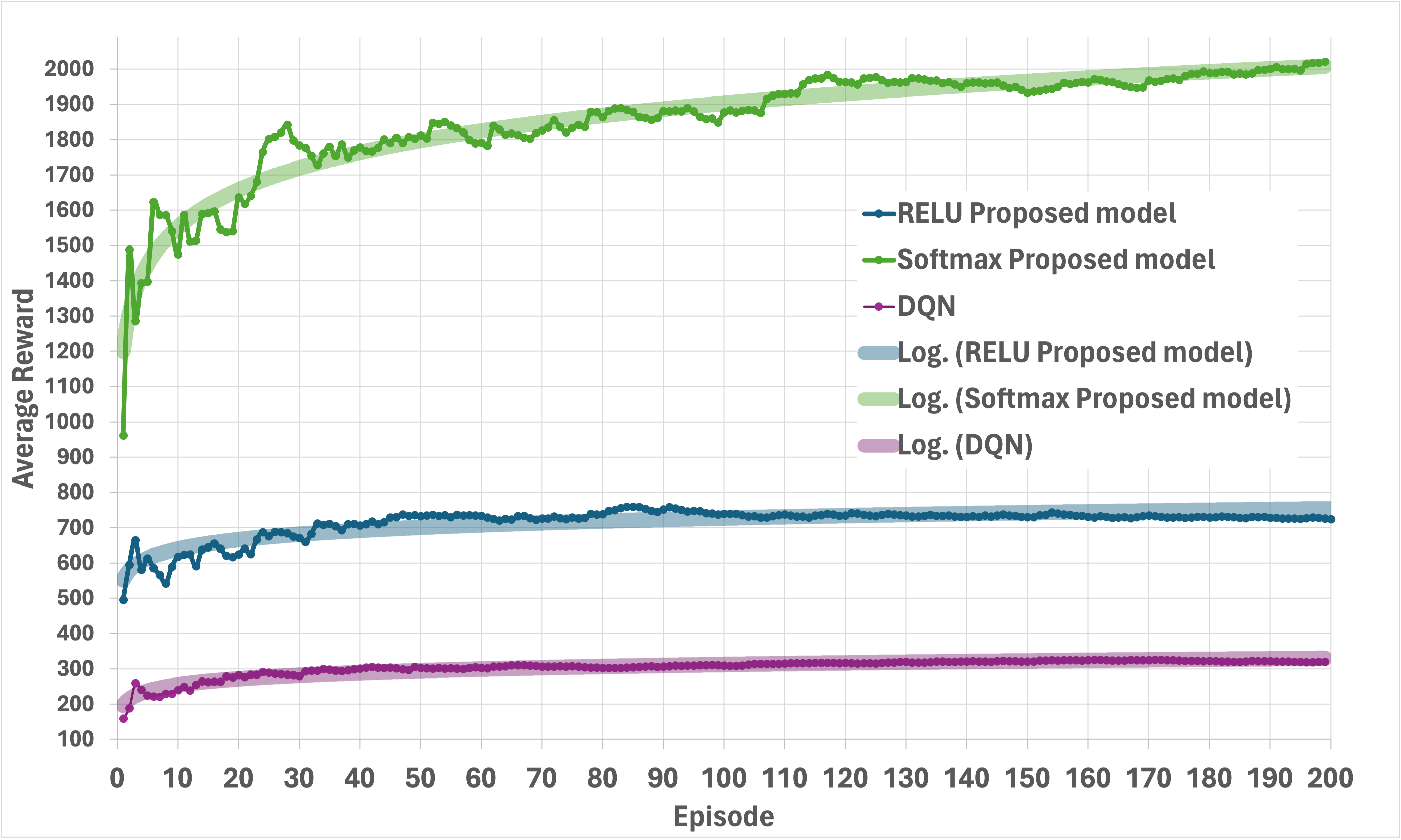}
     \caption{Convergence of the proposed MADDPG model with comparison to DQN algorithm}
     \label{fig:c}
    \includegraphics[width=0.9\linewidth]{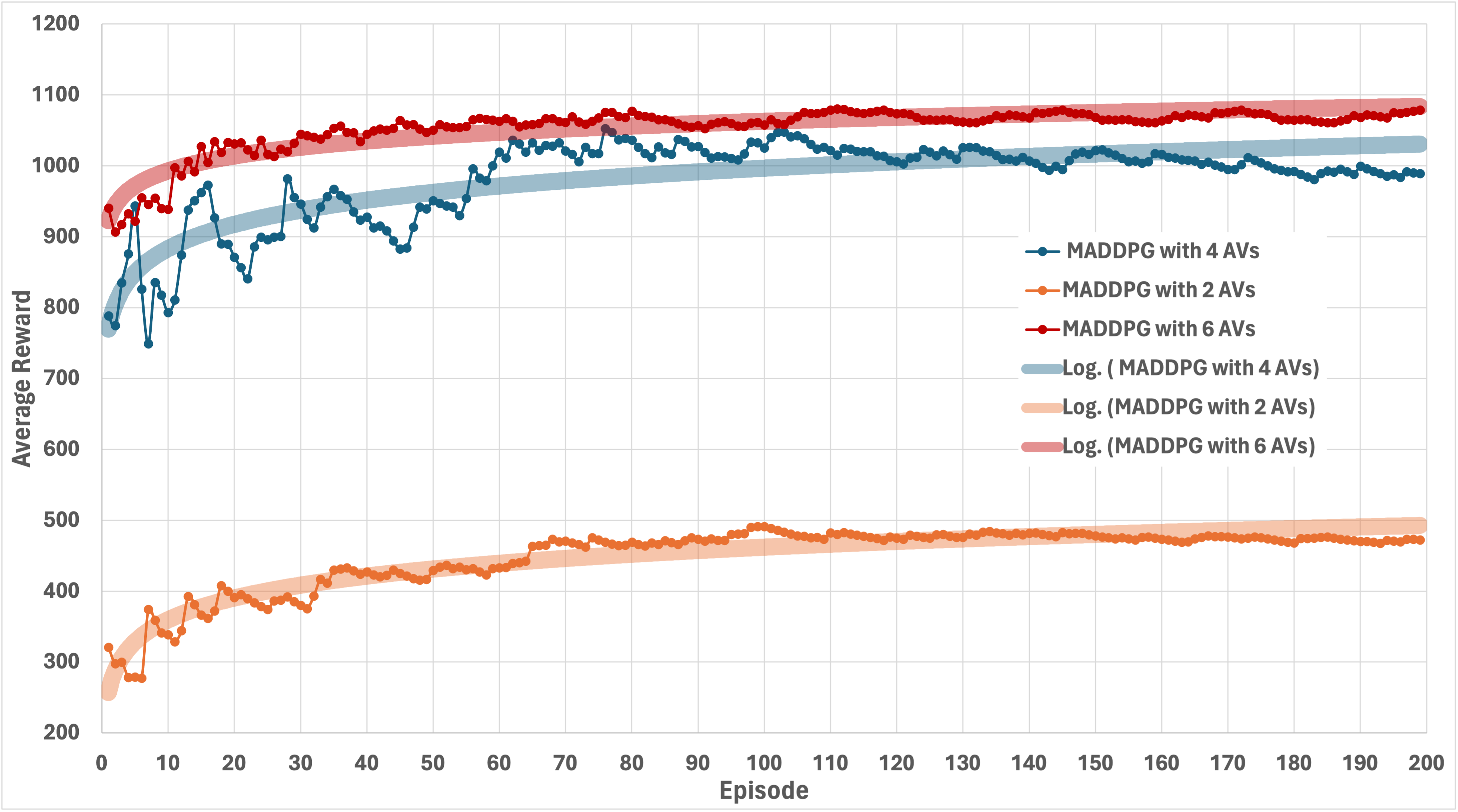}
     \caption{Scenario 1: 3-lane setting with two roadblocks evaluation for various number of vehicles.}
     \label{fig:b}
     \includegraphics[width=0.9\linewidth]{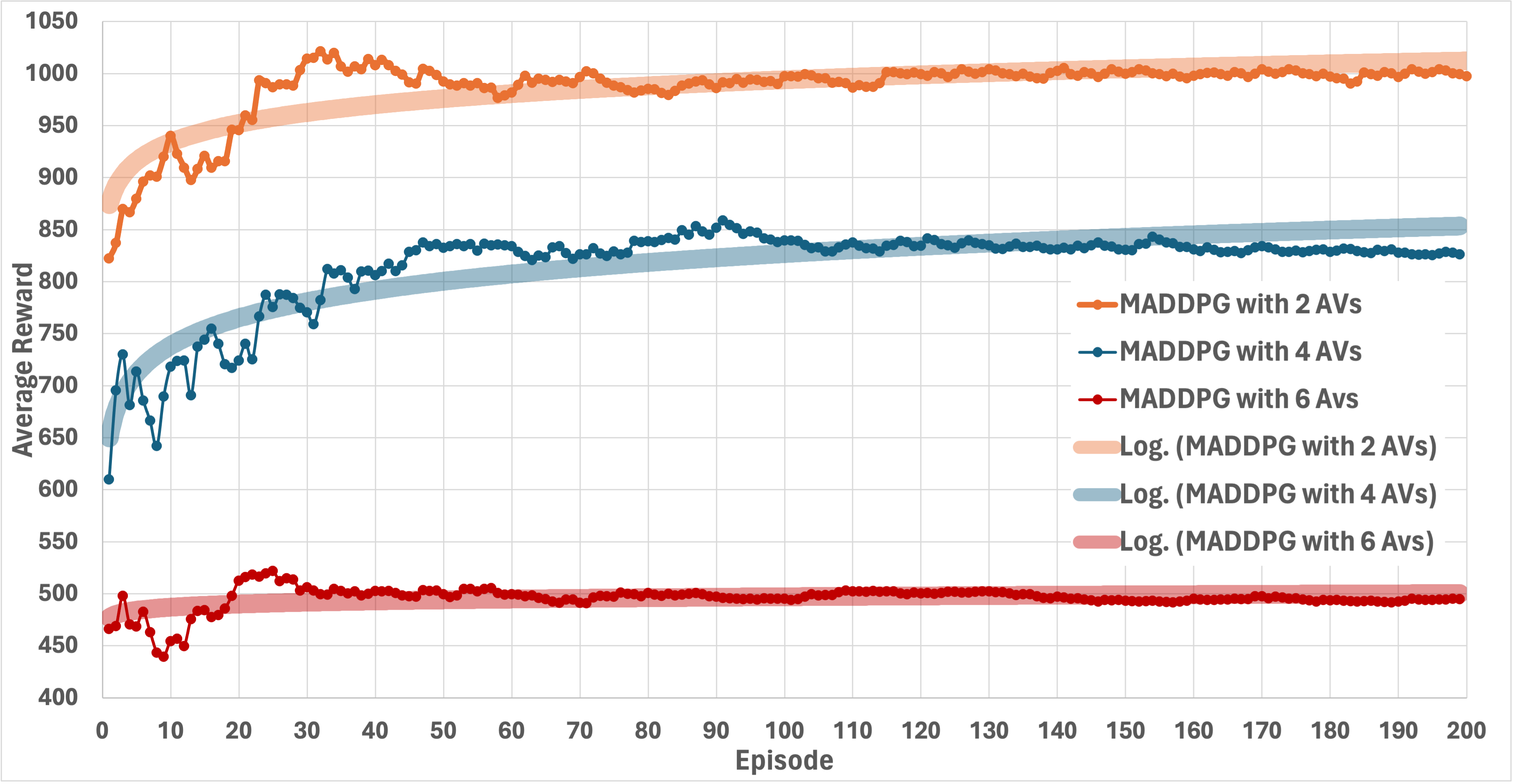}
     \caption{Scenario 2: 2-lane setting with one roadblock evaluation for various number of vehicles.}
     \label{fig:d}
\end{figure}

As shown in Fig.\ref{fig:c}, a comparison of the convergence of the proposed MADDPG algorithm with two activation networks and the DQN algorithm is demonstrated. MADDPG offers significant advantages over DQN, particularly in highly dynamic environments. Firstly, MADDPG employs actor-critic methods where the actor-network directly maps states to specific actions, and the critic-network evaluates the actions taken. This approach allows for more efficient and stable learning in highly variable environments. Secondly, MADDPG is designed specifically for multi-agent systems, enabling agents to learn policies that account for the actions and states of other agents. This is crucial in ITS, where vehicles need to coordinate to avoid collisions and optimize traffic flow. Thirdly, MADDPG provides more stable and sample-efficient training compared to DQN due to its use of deterministic policy gradients, which is particularly beneficial in dynamic environments where rapid adaptation is necessary. Lastly, MADDPG leverages centralized training with decentralized execution, allowing agents to access global information during training and act based on local observations during execution. This approach enhances scalability and practicality for real-world applications, making MADDPG more effective in multi-agent contexts than DQN.

The provided figures illustrate the performance of the proposed MADDPG  algorithm with varying numbers of AVs across different scenarios and metrics. As shown in Fig.\ref{fig:b}, the performance is shown over 200 episodes in a 2-lane scenario. The data reveals that the configuration with 6 AVs consistently achieves the highest average reward, stabilizing around 1100 after 100 episodes. In contrast, the configuration with 4 AVs shows moderate performance with stabilization around 900, and the configuration with 2 AVs performs the lowest, stabilizing around 500. This suggests that increasing the number of AVs enhances the performance and stability of the MADDPG algorithm in the 2-lane scenario. Meanwhile, in Fig.\ref{fig:d},  the performance of the MADDPG algorithm in a 3-lane scenario over 200 episodes is illustrated. Unlike the 2-lane scenario, the configuration with 2 AVs achieves the highest average reward, stabilizing around 1000. The 4 AVs configuration follows with an average reward stabilizing around 800, while the 6 AVs configuration performs the worst, stabilizing around 500. This indicates that in the 3-lane scenario, the performance decreases with the increment in the number of AVs, suggesting that the added complexity with more AVs negatively impacts the algorithm's effectiveness.

\begin{figure}
  \centering
  \captionsetup{belowskip=0pt}
     \includegraphics[width=0.9\linewidth]{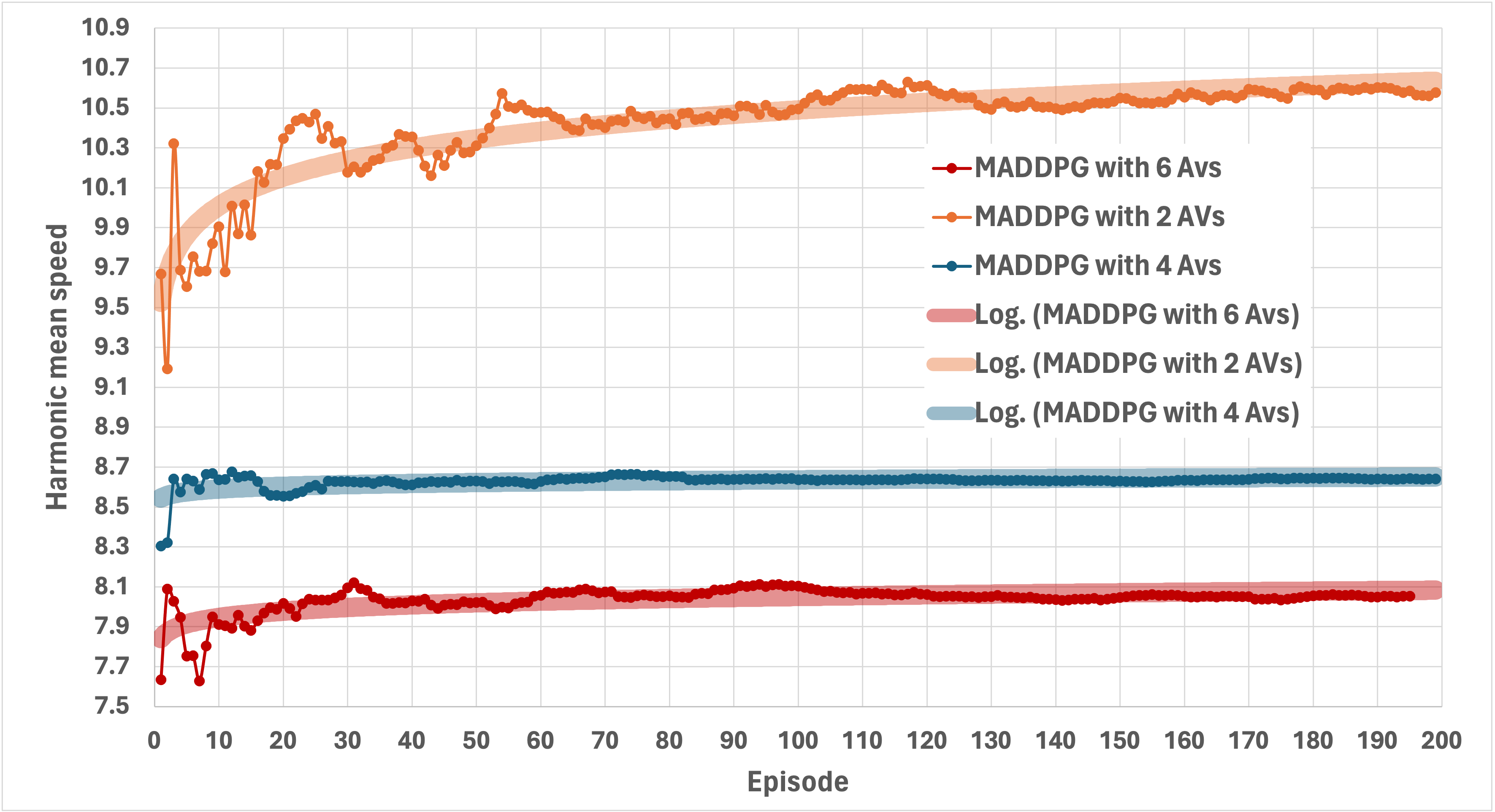}
     \caption{Harmonic mean speed evaluation for a 3-lane setting}
     \label{fig:g}
      \includegraphics[width=0.9\linewidth]{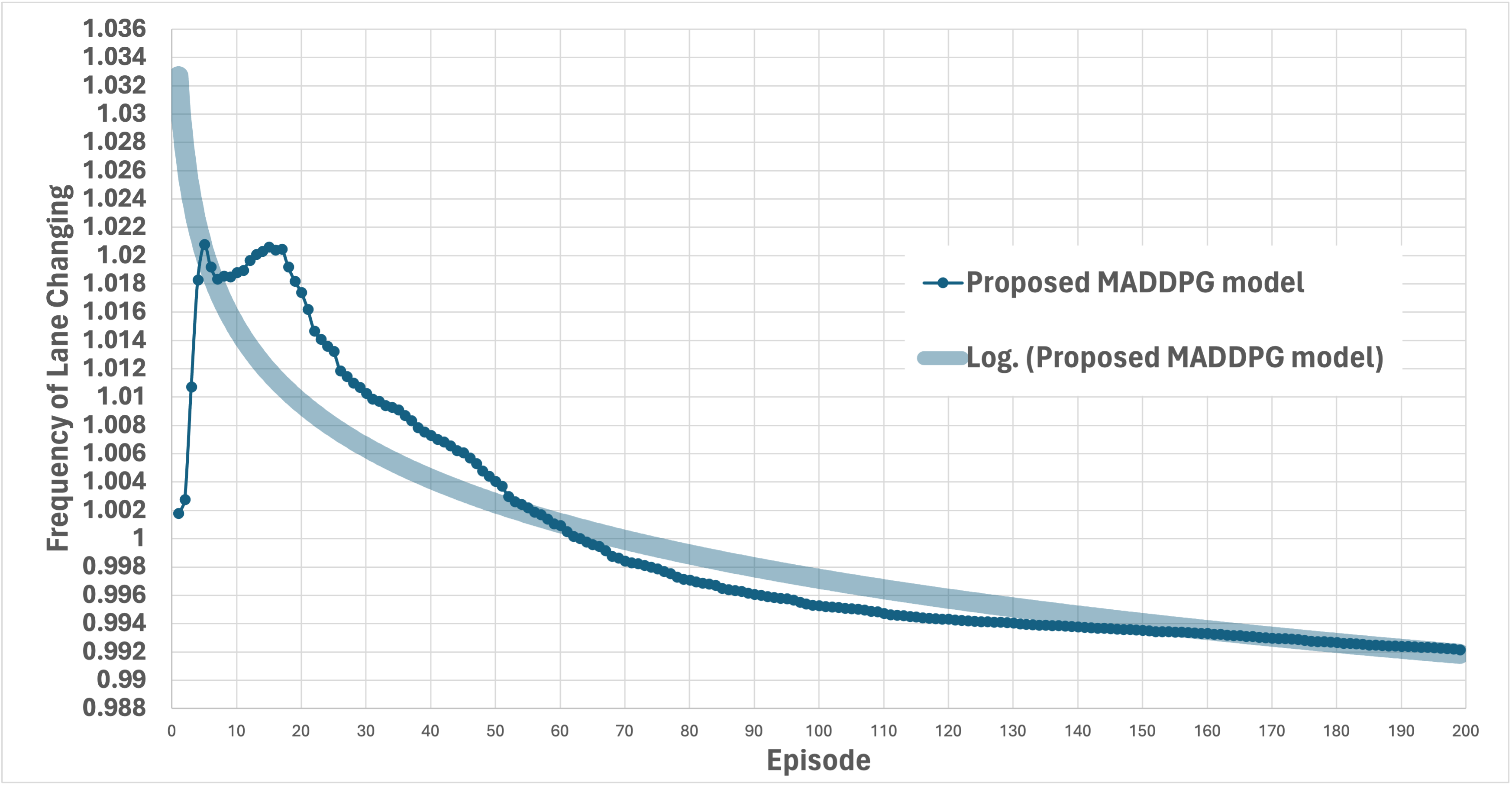}
     \caption{Lane changing frequency evaluation for a 3-lane setting}
     \label{fig:h}
\end{figure}

Finally, we evaluate the algorithm performance based on the harmonic mean speed as shown in Fig.~\ref{fig:g}. The figure displays the harmonic mean speed over 200 episodes for the different configurations of AVs. Here, the 2 AVs configuration achieves the highest harmonic mean speed, stabilizing around 10.4 $m/s$. The 4 AVs configuration stabilizes around 8.6 $m/s$, and the 6 AVs configuration has the lowest harmonic mean speed, stabilizing around 8.0 $m/s$. This trend shows that traffic flow efficiency, as measured by harmonic mean speed, is higher with fewer AVs, possibly due to reduced vehicle interaction complexity. When utilizing MADDPG to enhance traffic flow and avoid roadblocks in ITS, the observed decrease in lane-changing frequency over time can be attributed to the algorithm's ability to minimize unnecessary movements and improve vehicle coordination, as shown in Fig.~\ref{fig:h}. Frequent lane changes often lead to disruptions in traffic flow, causing stop-and-go waves and increasing the likelihood of congestion and accidents. our proposed algorithm helped agents learn to minimize such unnecessary movements by focusing on long-term rewards, which prioritize maintaining steady speeds, roadblock avoidance, and reducing abrupt lane changes. This results in a more stable and efficient traffic flow. Additionally, the algorithms enabled agents to learn from the actions and policies of other vehicles, fostering better anticipation of surrounding vehicles' behavior and reducing the need for reactive lane changes. The model encouraged agents to maintain lane discipline by optimizing traffic flow and reducing congestion, leading to a smoother and more predictable driving environment.

\section{Conclusion}

In this work, we presented a MARL framework designed to enhance AV decision-making within dynamic and complex ITS. Our primary goal is to enable AVs to avoid multiple dynamic roadblocks efficiently maintain optimal traffic flow, and maximize the mean harmonic speed under varying traffic conditions. Through the integration of key constraints such as minimum vehicle speed, roadblock count, and lane change frequency, we ensure realistic and practical operational scenarios. The evaluation of our proposed model has demonstrated significant improvements in traffic flow efficiency and robustness compared to other benchmark solutions such as DQN. Extensive simulations using the SUMO and TraCI interface confirmed the model's adaptability to diverse traffic conditions, effective navigation of dynamic roadblocks, and overall enhancement of AV operations. This work can be extended to incorporate advanced safety measures and robustness checks within the MARL framework to ensure fail-safe operations even in the presence of unexpected road conditions or sensor malfunctions.  

\bibliographystyle{IEEEtran}
\bibliography{references}
\end{document}